# Unsupervised Image Classification Through Time-Multiplexed Photonic Multi-Layer Spiking Convolutional Neural Network

M. Skontranis[(1)], G. Sarantoglou[(1)], S. Deligiannidis[(2)], A. Bogris[(2)], C. Mesaritakis[(1)]

[(1)] University of the Aegean, Dept. of Information and Communication Systems Engineering, Palama 2, Samos-83200 Greece mskontranis@icsd.aegean.gr
[(2)] University of West Attica, Dept. Informatics and Computer Engineering, Ag. Spiridonos, Egaleo, 12243-Greece.

**Abstract** *We present results of a deep photonic spiking convolutional neural network, based on two-section VCSELs, targeting image classification. Training is based on unsupervised spike-timing dependent plasticity, whereas neuron time-multiplexing and ultra-fast response are exploited towards a a reduction of the physical neuron count by 90%.*

**Introduction**

Spiking Neural Networks (SNNs) have risen as a promising alternative computational paradigm targeting machine-learning related problems. SNNs exhibit significant advantages such as information representation sparsity and low power consumption, which are based on their biologically inherited operation[1]. In order to harvest their merits, a critical aspect is the development of dedicated hardware platforms that can emulate neural operation and at the same time counterbalance electronics related limitations [2]. Towards this direction, Photonic Spiking Neural Networks (PSNN) based on laser neurons have drawn the spotlight of attention, due to high firing rate, low propagation losses, high wall-plug efficiency, isomorphism of dynamics to their biological counterparts and time/wavelength multiplexing capabilities[3].

Multitude of different photonic implementations have been studied such as two-section gain-absorber lasers, micro-disk lasers[3] and polarization switching VCSELS[4] etc. In particular VCSEL implementations hold very desirable attributes such as low power consumption and footprint thus allowing large – scale integration in 2D arrays. Recently a PSNN implementing a small – scale convolutional neural network (CNN) based on VCSEL neurons has been theoretically and experimentally implemented for image edge detection[5], as well as a PSNN that implements basic logical gate functions with polarization switching VCSELs under a supervised learning scheme[6]. Finally, an experimental implementation of a photonic retina-gaglion cell has been also achieved, taking advantage of excitatory and inhibitory dynamics[7].

In this work, we present simulation results from a fully deployed adaptation of a "deep" four-layer PSCNN, realized using two section VCSEL neurons, following the model developed in [8]. The proposed configuration exploits the low refractory period so as to time-multiplex inputs from different spatial locations in a Spatio-Temporal Encode Mapping (STEM) scheme. Therefore, the number of physical laser count can be drastically reduced by more than 90% compared to typical SCNNs[9]. Following this approach, processing speed can be tunable scaling from the multi-Gframe/sec regime using high neuron count to 120frame/sec rate equally reducing the number of neurons. The multi-layer time-multiplexed PSCNN scheme is evaluated by tackling an image processing task that consists of classifying noisy images of digits. Training in our case, contrary to previous approaches, is based on unsupervised photonic spike-timing dependent plasticity (STDP)[10], thus alleviating the need for complex offline processing.

**PSCNN's Structure and Function**

The proposed network architecture is a photonic adaptation of [9] (Fig. 1). The first layer encodes the magnitude of pixel's contrast to the latencies of the generated spikes. In the second layer, each neuron aims at identifying a simple spatial pattern, whereas the responses of the second layer are combined in the third and fourth layer, where the network detects more complex spatial patterns. A difference of our scheme compared to typical CNNs is the necessity for synchronizing inter-layer signals due to the time-multiplexing process. The Contrast detection layer (CDL) is designed to mimic the operation of the retina ganglion cell (RGC)[11]. Here, CDL implements a receptive field via a 3X3 layout of 9 neurons and the RGC itself via a single neuron (Fig. 2). Similar to the biological counterpart, the artificial receptive field is divided in two regions, namely the Center (C) (Fig. 2 dashed circle), implemented by a single excitatory VCSEL and the Surround (S) (Fig. 2 dotted circle), implemented by 8 inhibitory VCSELs. C and S neurons are biased at the excitable regime (a sufficient input triggers a spike) while the RGC is biased at the spiking regime (constantly firing spikes).[11] Using the aforementioned configuration, CDL encodes pixel's contrast in the spike's latency $t_x$; meaning that pixel's contrast is inverse proportional to the firing latency. More precisely the CDL scans images, pixel by pixel, using a scanning window of 3X3 pixels. The scanning window's center corresponds to the processing pixel of the image while the rest pixels of the scanning window are designated as surround pixels. Each pixel must be processed within a specific time-frame $T_{pr}$, due to time multiplexing. In terms of pixel's intensity, if the pixel is black then the input of the corresponding neuron is set to be an electrical rectangular pulse with constant power 0.2mW. Otherwise (white pixel) the input is set to 2μW.

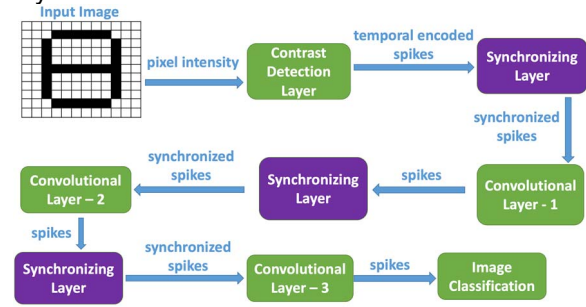

**Fig. 1** PSNN's architecture

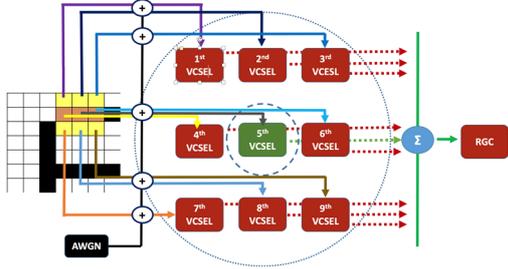

**Fig. 2** Structure of the photonic CDL, white noise (AWGN) is added at the VCSEL's current or at the mean pixel's intensity

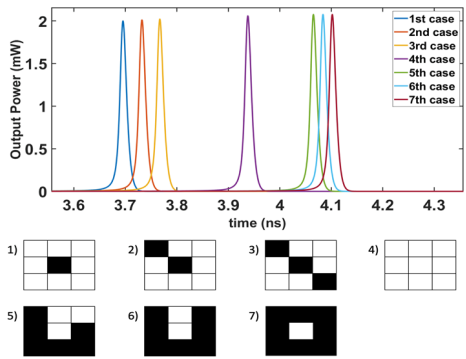

**Fig. 3** The temporal encoded output of the CDL for various spatial motifs

CDL's timing encoding scheme is based upon the excitatory and inhibitory dynamics of the C and S VCSELs. Due to spiking bias of the RGC VCSEL, a spike will always be produced at a specific time $t_x$. However, the input spike pattern alters $t_x$. Specifically, $t_x$ decreases if C-VCSEL is excited by a stimulus because the excitatory effect of C-VCSEL facilitates RGC's spiking. On the contrary, S-VCSELs hinders RGC's spiking upon activation causing the increase of $t_x$. Consequently, the number of black pixels and their spatial distribution determines $t_x$ delay. After the pixel's processing is completed, the process is repeated until the entire image is scanned. Contrary to [9], the output of the CDL here, is a spike train consisting of $N_{pix}$ spikes where $N_{pix}$ is the number of image's pixels. This occurs due to the time-multiplexing of spatial information, thus each spike is placed inside a predetermined time slot of period $T_{pr}$. Therefore, the $k_{th}$ spike encoding $k_{th}$ pixel's contrast will be in the $(k-1)T_{pr}$ - $kT_{pr}$ time slot. Nonetheless, from an operational point of view, these spikes should be simultaneously injected to the subsequent layer. Aiming to amend this, we introduce a synchronizing layer, so as spikes that belong to the same convolution window are delayed accordingly. It is important to mention that in our case we assume electro-optic coupling between neurons, thus long delays are easily implemented through the help of typical RF electronics.

Following the CDL, there is a Convolutional Layer (CL) whose task is to learn and detect the most frequent fundamental spike patterns (SP) that correspond to specific spatial features (contrasts). It comprises 40 VCSELs, each firing a spike only for a specific spatial feature. Its Convolutional Window ($CW_1$) is 3X3pixels. Each neuron has 9 inputs (for every pixel of the $CW_1$). The training of this layer is based on an unsupervised version of STDP [10]. More specifically, if the neuron under training (post-synaptic) fires a spike at $t_1$, then all spikes, originating from pre-synaptic neurons that arrived before $t_1$ will have their corresponding synaptic weights increased. On the contrary, spikes that arrive after $t_1$, will have their corresponding weight decreased. Using this principle, when random images are fed to the system each neuron will recognize a specific SP. Which neuron will lock is random, but STDP's weight update ensures that this neuron will re-fire when this SP emerges. Furthermore, in order to ensure that only a single neuron will be triggered for each SP, the activated VCSEL sends a cancelling signal to all of the following VCSELs in order to ignore this particular SP. The cancelling signal lowers the injection current bias of the following neurons, forcing them away from their excitable regime. This process continues until each SP is processed. Another crucial issue is that STDP will boost the synaptic weights of the neuron with the lowest spike latency [8], regardless if this neuron has already "learned" an SP. In order to hinder this behavior, the maximum and minimum weight values of the STDP are adjusted. This process is the operational equivalent of a maximum pooling layer [9] and setting the final weights requires no supervision, only a prior knowledge of the number of inputs for each neuron. The following CLs consist of 16 and 8 VCSELs while their CW comprise 6x6 and 6x12 pixels respectively. Their operating principle is similar to the first layer, except their received input; meaning that $CL_1$ does not produce temporal-encoded signals and firing activity occurs only if a specific SP is detected. Therefore, this "binary" input leads to a training method where the existence of a spike results to an increase of synaptic weight by a constant value independent of the time of arrival.

**Application Scenario**
The aforementioned PSCNN was realized using C++ and simulations were performed by Nvidia's Titan RTX GPU. The training data consisted of four 12X12 pixel, black and white images (Fig. 4a). During inference we enriched our data set using two approaches; adding white noise to the driving currents of the VCSELs with signal to noise ratio (SNR) down to 3dB and using images whose pixel's intensity was perturbed by adding a random offset. In $CL_1$ simple patterns are identified (Fig. 4b). The $CW_1$ is shifted one pixel at a time in the horizontal/vertical direction, allowing the scanning of each image. In $CL_2$ the network learns to detect more complex patterns, which are combinations of $CL_1$'s motifs (Fig. 4c). The shift of $CW_2$, horizontally and vertically, is equal to pixels per row of $CW_2$ (6 pixels). The same principles apply to $CL_3$, where its patterns are combinations of $CL_2$'s (Fig. 4d).

After STDP training, we performed inference for the two above mentioned cases. The first, where additive noise at the VCSEL's input current was used, generated no misclassification, despite the low SNR. We assume that this resilience to system noise stems from the integrating nature of the neurons and also relies on the noise cancelling effect of the spiking process [1]. In the second case, we have added random variations (following a normal distribution) to the intensity of each pixel (fig. 5 insets) and we have computed the classification error versus the standard deviation of the variation (fig. 5). It can be seen that up to std=0.06 no classification error can be computed, while an abrupt transition occurs for higher values. The origin of this sensitivity can be traced to the CDL that generates different SPs, due to the contrast variations at the image.

**Conclusion**
The proposed PSCNN consists of VCSELs in a "deep" architecture of four layers, whereas excitation and inhibition are realized through electro-optic modulation. Our configuration is versatile, as it can classify images using STDP in a purely unsupervised manner. Another key advantage of the proposed architecture is the radical reduction of physical neuron count, due to spatio-temporal multiplexing. In a typical SCNN the image is divided in several areas, known as neural maps [9]. Each neural map is assigned to a specific group of neurons, that are detecting SPs. Consequently, different neurons detect the same pattern but for a different neural map. In our scheme, one neuron is used to detect a specific SP for all neural maps leading to a 90% reduction of physical neurons, equally affecting the energy footprint of the scheme. This advantage fits really well to photonic implementations, where ultra-low response time is well within grasp, while dense integration remains partially elusive. Finally, the proposed scheme exhibits increased resilience to systematic noise (current, photodiodes etc.) while similar to typical SCNNs [12] the network is sensitive to image contrast variations due to the temporal encoding.

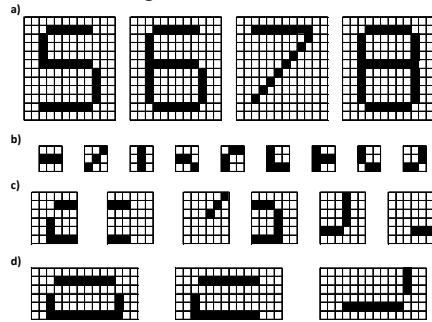

**Fig. 4** a) target images b) First, c) Second, d)Third CL trained SPs

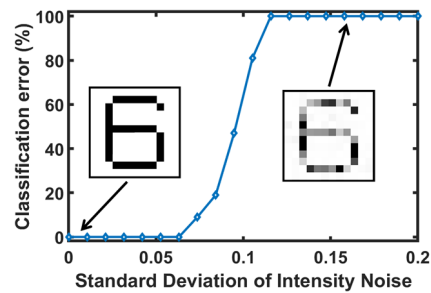

**Fig. 5** Classification error versus the standard deviation of the intensity purturbation added at each pixel.

**Acknowledge**
This project has received funding from the Hellenic Foundation for Research and Innovation (HFRI) and the General Secretariat for Research and Technology (GSRT), under grant agreement No 2247 (NEBULA project)